\documentclass[a4paper,fleqn]{cas-dc}

\usepackage[mathscr]{eucal}
\usepackage{amsmath}
\usepackage{amssymb}
\usepackage{mathtools}
\usepackage{amsthm}
\usepackage{graphicx} 
\usepackage[utf8]{inputenc}
\usepackage[english]{babel}
\usepackage[svgnames]{xcolor}
\usepackage[svgnames]{xcolor}
\usepackage[numbers]{natbib}
\usepackage{setspace}

\theoremstyle{remark}
\newtheorem{remark}{Remark}
\newtheorem{proposition}{Proposition}

\renewcommand{\=}{\stackrel{\mbox{\scriptsize def}}{=}}
\newcommand{\I}{\text{i}}
\renewcommand{\d}{\text{d}}

\renewcommand\Re{\operatorname{Re}}
\renewcommand\Im{\operatorname{Im}}

\newcommand{\A}{a}
\newcommand{\Vv}{v}
\newcommand{\phii}{{u}}

\begin{document}
\selectlanguage{english}

\let\WriteBookmarks\relax
\def\floatpagepagefraction{1}
\def\textpagefraction{.001}
\shorttitle{Energy transport in a free Euler-Bernoulli beam in terms of Schrödinger's wave function}
\shortauthors{Serge~N. Gavrilov et~al.}

\title [mode = title]{Energy transport in a free Euler-Bernoulli
beam in terms of Schrödinger's wave function}
\tnotemark[1]
\tnotetext[1]%
{The manuscript of this paper is available as \href{https://doi.org/10.48550/arXiv.2411.04033}{arXiv preprint 2411.04033}
in full accordance with \href{https://www.elsevier.com/about/policies-and-standards/sharing}{Elsevier Article Sharing Policy}.}
%
%

\author%
[1]{Serge~N. Gavrilov}[orcid=0000-0002-7889-3350]
\ead{serge@pdmi.ras.ru}
\cormark[1]
\credit{Methodology, Formal analysis, Writing --- original draft, Writing --- review {\&} editing}
\author%
[1,2]{\,Anton~M. Krivtsov}[orcid=0000-0002-9258-065X]
\ead{akrivtsov@bk.ru}
\credit{Conceptualization, Methodology, Formal analysis, Writing --- review {\&} editing}

\author%
[1]{\,Ekaterina~V. Shishkina}[orcid=0000-0003-4829-9403]
\ead{shishkina_k@mail.ru}
\credit{Methodology, Formal analysis, Writing --- review {\&} editing}

\address[1]{Institute for Problems in Mechanical Engineering RAS, St.~Petersburg, Russia}
\address[2]{Peter the Great St.~Petersburg Polytechnic University (SPbPU), St.~Petersburg, Russia}

%
%
%
%
%
\cortext[cor1]{Corresponding author}
%




\maketitle

\begin{abstract}
 The Schrödinger equation is not frequently used in the framework of the
 classical mechanics, though historically this equation
 was derived as a simplified equation, which is equivalent to
 the classical Germain-Lagrange dynamic plate equation. 
 The question concerning the exact meaning of this equivalence is still
 discussed in
 modern literature. In this note, we consider the one-dimensional case, where the
 Germain-Lagrange equation reduces to the Euler-Bernoulli equation, which
 is used in the classical theory of a beam. We establish a one-to-one correspondence
 between the set of all solutions (i.e., wave functions $\psi$)
 of the 1D time-dependent Schrödinger equation for a
 free particle with arbitrary complex valued initial data and the set of ordered pairs
 of quantities (the beam strain and the particle velocity),
 which characterize solutions $\phii$ of the beam equation with arbitrary real
 valued initial data. Thus, the dynamics of a free infinite Euler-Bernoulli beam
 can be described by the Schrödinger equation for a free particle and vice
 versa. 
 Finally, we show that
 for two corresponding solutions $\phii$ and $\psi$ the mechanical energy
 density calculated for $\phii$ propagates in the beam exactly in the same way
 as the probability density calculated for $\psi$.
\end{abstract}

\section{Introduction}
 In the framework of the classical mechanics, the Schrödinger equation
 commonly can be found only while solving some particular problems.
 For example, the Schrödinger-type equation naturally
 emerges when constructing localized solutions of 3D and 2D wave equation, see,
 e.g., review \cite{Kiselev2007} and corresponding references
 therein.\footnote{Though, the wave equation is more often discussed in
 the context of optics rather than mechanics.} Analogously, this equation can
 appear when investigating some elastodynamics problems
 \cite{Perel2005}.
 However, historically, the Schrödinger equation was introduced in
 \cite{SchroedingerAdP4-1926,Schroedinger1982}
 as a simplified equation, which, according to \cite{SchroedingerAdP4-1926}, 
 is equivalent to the classical Germain-Lagrange dynamic plate equation of
 fourth-order in space and second-order in time with some additional terms
 related to the external potential. Schrödinger stated  that 
 these two equations are equivalent, and this issue is still discussed in 
 modern
 literature \cite{Engstrom2023,Makris2023,Volovich2024}. In \cite{Engstrom2023} a 1D
 quantum system is corresponded to an equivalent mechanical system, which
 consists of two coupled beams. 
 In recent paper by Makris \cite{Makris2023}, 1D time-independent Schrödinger 
 equation with non-zero external potential
 is considered in the pure quantum framework. The solutions for spectral
 problems for the original, fourth-order in space equation, and the Schrödinger
 equation in the modern form 
 are compared.
 Note that the reduction of the original equation 
 to the Schrödinger equation in its modern form in \cite{SchroedingerAdP4-1926}
 is a bit similar to the mathematical approach
 previously applied by E.~Schrödinger 
 while investigating a completely different problem in the framework of the classical mechanics.
 This was the problem concerning the motion of 1D chain of material 
 particles \cite{schrodinger1914dynamik,Muehlich2020}.

 In this note, we also discuss the equivalence and restrict ourselves 
 to the simplest one-dimensional case,
 where the Germain-Lagrange equation reduces to the 1D Euler-Bernoulli
 equation, which is used in the classical theory of a beam. We restrict
 ourselves with the 1D time-dependent
 Schrödinger equation for a free particle only, i.e., the case
 when an external potential is zero. This equation has the form of:
\begin{equation}
\I \hbar \dot{\varPsi}+\frac{\hbar^2}{2m}{\varPsi''}=0, 
\label{Schr-eq-orig}
\end{equation}
and can be rewritten as follows:
\begin{gather}
\mathcal S_+ \varPsi
=\I a \dot{\varPsi}+b{\varPsi''}=0, 
\label{Schr-eq}
\\
\mathcal S_+\=\I a\frac{\partial }{\partial t}+ b \frac{\partial^2 }{\partial x^2},
\end{gather}
see classical textbook \cite{Messiah1-1961}. 
Here prime
denotes the derivative with respect to a spatial co-ordinate $x$, overdot is the
derivative with respect to time $t$.
In the quantum framework,
\begin{equation}
  a=\hbar=\frac{h}{2\pi}>0, \qquad 
  b=\frac{\hbar^2}{2m}>0,
\end{equation}
where
$\hbar$ is the reduced Planck constant, $h$ is the Planck constant,
$m$ is the particle mass, $\varPsi(t,x)$ is the wave function.
Multiplying Eq.~\eqref{Schr-eq} by the operator 
\begin{equation}
\mathcal S_-\=\mathcal S_+^\ast=-\I a\frac{\partial }{\partial t}+ b \frac{\partial^2 }{\partial
x^2}
\end{equation}
of the complex conjugate Schrödinger equation
\begin{gather}
\mathcal S_+^\ast \varPsi
= 
-\I a\dot{\varPsi}+b{\varPsi''}
=0
\label{Schr-eq-c}
\end{gather}
results in the Euler-Bernoulli dynamic beam equation: 
\begin{gather}
\mathcal B\varPsi=
a^2\ddot\varPsi
+b^2 \varPsi''''
=0,
\label{beam-eq-u}
\\
\mathcal B
\=\mathcal S_\pm\mathcal S_\mp
=\mathcal S_\pm\mathcal S_\pm^\ast
=
\A^2\frac{\partial^2 }{\partial t^2}+ b^2 \frac{\partial^4 }{\partial x^4}.
\end{gather}
Here, the asterisk symbol denotes the complex conjugation. In the framework of
the beam theory, $a^2$ is the mass of the beam per unit length, $b^2$ is the flexural rigidity,
$\varPsi(t,x)$ are the displacements, see, e.g., \cite{Weaver1990}.
One can see that any solution of Eq.~\eqref{Schr-eq} is the solution of
Eq.~\eqref{beam-eq-u}. Moreover, it is clear that, in principle, Eq.~\eqref{beam-eq-u} can be
solved by successive resolving of two conjugate Schrödinger-type equations
\cite{Orsingher2011,KorikovSch2021}. Note that in the quantum framework the
quantity $\varPsi$ is complex valued, whereas in the beam theory it is real
valued.

Discussing the system equivalence, we need to consider not only the
corresponding equations, but also to take into account the initial data.
In this note, 
we want to compare sets of all solutions of Eqs.~\eqref{Schr-eq}, 
\eqref{beam-eq-u}, respectively, which satisfy arbitrary initial data.
We establish a
one-to-one correspondence (bijection) between the set 
of all solutions 
of the Schrödinger equation~\eqref{Schr-eq} (wave-functions)
for a free particle with arbitrary complex valued initial
data, and the set 
of ordered pairs of quantities (the beam strain\footnote{We mean
the strain in the framework of the direct approach \cite{Altenbach2013,Zhilin2007}
to the beam theory, not the
strain in a beam considered as a three-dimensional elastic body.} and the
particle velocity), which characterize solutions 
of the beam equation~\eqref{beam-eq-u}
with arbitrary real valued initial data. Moreover, any initial value problem for a free
infinite Euler-Bernoulli beam with real valued initial data
can be reformulated in terms of an initial value problem formulated 
only for the Schrödinger equation \eqref{Schr-eq}, without any consideration
of complex conjugate equation \eqref{Schr-eq-c}. 
Thus, the order of the equation we need to solve can be reduced, at the cost
of we find only the strain and the particle velocity, not the
displacements. 
Finally, we show that for two
corresponding solutions of the beam equation and the Schrödinger equation, the mechanical energy density
calculated for the first solution propagates in the beam exactly in the same way as the
quantum probability density calculated for the second one.

\section{The initial value problem for the beam
equation, equivalent to a given initial value problem for the Schrödinger equation}
\label{S2}

Consider a classical initial value problem for the Schrödinger-type equation
for an unknown function $\varPsi_\pm(t,x)$:
\begin{gather}
\mathcal S_\pm \varPsi_\pm
\equiv 
\pm\A\I \dot{\varPsi}_\pm+b{\varPsi_\pm''}
=0,
\label{Schr-eq-chi}
\\
\varPsi_\pm(0,x)=\varPsi_\pm^0(x).
\label{ic}
\end{gather}
Here ${\varPsi_\pm^0}(x)$ is a given arbitrary complex valued function of $x$.
Equation~\eqref{Schr-eq-chi} is formulated in domain $t>0,\ x\in\mathbb R$.
This problem can be equivalently reformulated \cite{Vladimirov1971} as the
following initial value problem for an unknown distribution (or a generalized
function) $\varPsi_\pm$:
\begin{gather}
\mathcal S_\pm \varPsi_\pm=\pm\A\I\varPsi_\pm^0(x)\delta(t),
\label{Schr-eq-ic}
\\
\varPsi_\pm(t,x)\big|_{t<0}\equiv0,
\label{psi-is0}
\end{gather}
where
$\delta(t)$ is the Dirac delta-function.
Equation~\eqref{Schr-eq-ic} is formulated in domain $t\in\mathbb R,\ x\in\mathbb R$.
\begin{remark} 
  Everywhere in this note, discussing  ``arbitrary'' initial conditions, we mean
  initial conditions such that the solution of the corresponding initial value
  problem exists in the sense of the generalized functions. Moreover, we
  consider two initial value problems to be equivalent, if their solutions are always 
  equal to each other, provided that both solutions exist. The question concerning the existence of the 
  corresponding solutions is beyond the scope of this note.
\end{remark} 
Multiplying Eq.~\eqref{Schr-eq-ic} by 
$\mathcal S_\mp=\mathcal S_\pm^\ast$
results in
\begin{gather}
\mathcal B\varPsi_\pm
=
\A^2\varPsi_\pm^0(x)\dot\delta(t)\pm\I \A b {\varPsi_\pm^0}''(x)\delta(t).
\label{beam-eq-v-ic}
\end{gather}
For a smooth enough 
function $\varPsi_\pm^0(x)$, the generalized initial value problem in the form of 
Eqs.~\eqref{psi-is0}, \eqref{beam-eq-v-ic}
is equivalent to the classical initial value problem for the
homogeneous Euler-Bernoulli beam equation
\begin{gather}
\mathcal B
\varPsi_\pm
=0
\label{beam-eq-u-phi}
\end{gather}
with the following initial conditions
\begin{equation}
\varPsi_\pm(0,x)=\varPsi_\pm^0(x),\qquad \dot\varPsi_\pm(0,x)=\pm \frac {\I b}\A{\varPsi_\pm^0}''(x).
\label{Sro-gen-ic}
\end{equation}
Since the generalized solution of the initial value problem is unique, we have
proved the following proposition:
\begin{proposition}
  \label{prop:S}
  Generalized initial value problem
\eqref{Schr-eq-ic},
\eqref{psi-is0}
  for the Schrödinger-type equation 
  is equivalent to the generalized initial value problem
\eqref{beam-eq-v-ic},
\eqref{psi-is0}
for the Euler-Bernoulli equation.
The corresponding classical initial value problems are
expressed by Eqs.~\eqref{Schr-eq-chi},
\eqref{ic}
and 
Eqs.~\eqref{beam-eq-u-phi}, \eqref{Sro-gen-ic},
respectively.
\end{proposition} 

\begin{remark}
Since generally $\varPsi_\pm^0(x)$ is complex valued, the corresponding initial data for the Euler-Bernoulli equation 
\eqref{beam-eq-u-phi}
are also complex valued.
\end{remark} 

Proposition~\ref{prop:S} has been obtained independently from us in recent preprint \cite{Volovich2024}.

\section{The initial value problem for the Schrödinger equation, equivalent to
a given initial value problem for the beam equation}
\label{S3}
\subsection{The case of complex valued initial data for the beam equation}

Consider now a general classical initial value problem for the Euler-Bernoulli equation
\begin{gather}
\mathcal B
\phii
=0
\label{beam-gen}
\end{gather}
with the following initial condition:
\begin{gather}
\phii(0,x)=\phii^0(x),
\qquad
\dot\phii(0,x)=\dot\phii^0(x).
\label{ic-phi-again}
\end{gather}
Here $t>0,\ x\in\mathbb R$; ${\phii}^0(x)$, ${\dot\phii^0}(x)$ are given arbitrary complex valued functions of $x$.
This is equivalent to the following generalized initial value problem:
\begin{gather}
\mathcal B\phii
=
\A^2\phii^0(x)\dot\delta(t)+\A^2\dot\phii^0(x)\delta(t),
  \label{eq:beam-delta1}
\\
\phii\big|_{t<0}=0,
  \label{eq:beam-delta2}
\end{gather}
where $t\in\mathbb R,\ x\in \mathbb R$.

We want to represent the solution of the generalized initial value problem 
\eqref{eq:beam-delta1}, \eqref{eq:beam-delta2} in the form of a superposition 
\begin{equation}
  \phii=\varPsi_++\varPsi_-,
\label{super}
\end{equation}
where $\varPsi_\pm(t,x)$ satisfy Eqs.~\eqref{Schr-eq-ic}, \eqref{psi-is0}.
Substituting Eq.~\eqref{super} into Eq.~\eqref{eq:beam-delta1} and expressing
$\mathcal B\varPsi_\pm$ by virtue of Eq.~\eqref{beam-eq-v-ic}, one can equate
the terms proportional to $\delta(t)$ and $\dot\delta(t)$. This yields:
\begin{gather}	
{\varPsi_+^0}+{\varPsi_-^0}={\phii^0},
\label{EB-1st-ic}
\\
{\I b}
\big({\varPsi_+^0}''-{\varPsi_-^0}''\big)
=\A\dot{\phii}^0.
\label{EB-2nd}
\end{gather}
Differentiating Eq.~\eqref{EB-1st-ic} with respect to $x$ twice, one gets
\begin{gather}	
  {\varPsi_+^0}''+{\varPsi_-^0}''={\phii^0}''.
\label{EB-1st-ic-mod}
\end{gather}
Resolving the set of equations \eqref{EB-2nd},
\eqref{EB-1st-ic-mod} yields 
\begin{gather}
{\varPsi_\pm^0}''
=\frac1{2b}\big(\mp\I\A\dot{\phii}^0+{b}{\phii^0}''\big).
\label{psi12-1}
\end{gather}
Then functions $\varPsi_\pm^0(x)$ in the form of the equation
\begin{gather}
{\varPsi_\pm^0}
=\frac1{2b}\iint\big(\mp\I\A\dot{\phii}^0+b{\phii^0}''\big)\,\d x+ C_\pm x+ D_\pm
\label{chi-int}
\end{gather}
are the initial data for
functions $\varPsi_+$ and $\varPsi_-$ introduced by Eq.~\eqref{super}. 
Here, $C_\pm$ and $D_\pm$ are arbitrary constants.
One can take:
\begin{equation}
  C_\pm=0,\qquad D_\pm=0.
  \label{eq:CD0}
\end{equation}

Thus, according to Proposition~\ref{prop:S}, one has:
\begin{proposition} 
  \label{prop:S2}
The solution $\phii$ of initial value problem
\eqref{beam-gen},
\eqref{ic-phi-again}
for the Euler-Bernoulli equation
with arbitrary smooth complex valued initial data $\phii^0$, $\dot\phii^0$
can be found in the form of the superposition \eqref{super},
where 
${\varPsi_\pm}(t,x)$ 
are the solutions of 
initial value problems for two complex conjugate Schrödinger-type equations~\eqref{Schr-eq-chi} with 
initial conditions \eqref{ic},
wherein ${\varPsi_\pm^0}(x)$ 
are found by Eq.~\eqref{chi-int}.
\end{proposition} 



\subsection{The case of real valued initial data for the beam equation}
\label{S3-1}
Let $\phii^0(x),\ \dot\phii^0(x)$ be real valued for all $x$. 
In this case, according to Eqs.~\eqref{chi-int}, \eqref{eq:CD0}
\begin{equation}
\varPsi_+^0(x)
=
\big(\varPsi_-^0(x)\big)^\ast,
\end{equation}
and, thus,
\begin{gather}
\varPsi_+(t,x)
=
\varPsi_-^\ast(t,x) 
\end{gather}
for all $t\geq0$, and
\begin{gather}
\phii(t,x)=\varPsi_++\varPsi_+^\ast=2\Re\varPsi_+(t,x),
\label{pchi1}
\\
\dot\phii(t,x)=\frac{\I b}{\A}\big(\varPsi_+''-(\varPsi_+^\ast)''\big)=-\frac{2 b}{\A}\Im\varPsi_+''(t,x),
\label{pchi2}
\end{gather}
due to Eqs.~\eqref{super}--\eqref{EB-2nd}.
Now, Proposition~\ref{prop:S2} can be reformulated as follows:
\begin{proposition} 
  \label{prop:S3}
The solution $\phii$ of initial value problem
\eqref{beam-gen},
\eqref{ic-phi-again}
for the Euler-Bernoulli equation
with arbitrary smooth real valued initial data $\phii^0$, $\dot\phii^0$
can be found by formula~\eqref{pchi1}. Here
${\varPsi_+}(t,x)$ 
is the solution of the initial value problem 
for the Schrödinger equation \eqref{Schr-eq-chi} with initial condition \eqref{ic},
wherein ${\varPsi_+^0}(x)$ is found by Eq.~\eqref{chi-int}.
\end{proposition} 

\begin{remark} 
  Thus, we have established the correspondence $\phii\rightarrow\varPsi_+$ between 
\begin{itemize}
  \item
  the set 
  of all solutions $\phii$ (displacements) of the beam equation
  with arbitrary real valued initial data;
  \item 
  the set 
  of solutions $\varPsi_+$ (wave-functions) of the Schrödinger
  equation~\eqref{Schr-eq-chi}
  with complex valued initial data.
\end{itemize}
However, such a correspondence $\phii\rightarrow\varPsi_+$ is not useful in practice since double
integration is required to find the initial data $\varPsi_+^0$ for given
$\phii^0$. Moreover, trying to generalize this result to 2D and 3D cases, we
get $\nabla^2$ instead of double differentiation in the left-hand side of Eq.~\eqref{psi12-1}. 
Hence, double integration in 
Eq.~\eqref{chi-int}
transforms into solving of the Poisson equation. In what follows, we suggest
another correspondence, which is free from this drawback.
\end{remark} 



Introduce now the following complex quantity:
\begin{equation}
\psi(t,x)\=-\I\A\dot{\phii}+b{\phii}''=2b\varPsi_+''(t,x),
\label{psi-via-chi1}
\end{equation}
where the last equality is due to Eqs.~\eqref{pchi1}, \eqref{pchi2}.
Function
${\psi}(t,x)$ satisfies
the Schrödinger equation
\begin{equation}
\mathcal S_+\psi=0
\label{Spsi}
\end{equation}
and initial conditions, which correspond to Eq.~\eqref{psi-via-chi1} taken at
$t=0$.
According to
Eq.~\eqref{psi-via-chi1},
one has 
\begin{gather}
\psi=b\gamma\{\phii\}-\I\A \Vv\{\phii\},
\label{icPsi}
\\
  \gamma
  (t,x)\=
\phii''(t,x),
\label{Ppsi1}
\\
\Vv
(t,x)\=
\dot\phii(t,x).
\label{Ppsi2}
\end{gather}
Here 
$\Vv\{\phii\}$ is the particle velocity,
and $\gamma\{\phii\}$ is the {beam strain}, 
which corresponds to the displacement $\phii$.

\subsection{Energy transport}
Assuming that $\phii$ is a real valued function,
one can obtain the equation for the balance of energy in the beam by means of
multiplying Eq.~\eqref{beam-gen} by $v$. After some transformations, we
get
\begin{equation}
\dot{\mathcal E}=-Q', 
\label{e-balance}
\end{equation}
where 
\begin{gather}
  \mathcal
  E\{\phii\}=\frac
  {b^2}2\big(\phii''\big)^2+ \frac{\A^2}{2}\big(\dot\phii\big)^2,
  \label{E-BE}
  \\
  Q\{\phii\}=b^2(v\gamma'-v'\gamma)
\end{gather}
are the mechanical energy density and the mechanical energy flux, 
respectively,
which correspond to the solution $\phii$ of the Euler-Bernoulli equation.
According to  Eq.~\eqref{e-balance},
\begin{equation}
  \frac\d{\d t} \left(\int_{-\infty}^{+\infty}\mathcal E(t,x)\,\d x\right)=0
  \label{conserves}
\end{equation}
provided that $Q\to 0$ as $x\to\infty$, i.e., the energy conserves.
Let us calculate the quantum 
probability density $\rho\{\psi\}$ \cite{Messiah1-1961}
\begin{equation}
\rho
\=\lambda|\psi|^2=\lambda{\psi\psi^\ast},
\label{E-S}
\end{equation}
which corresponds to a wave-function $\psi$
defined by Eq.~\eqref{icPsi}. 
According to Eqs.~\eqref{icPsi}--\eqref{Ppsi2},
we have
\begin{gather}
  \rho
  =
  2\lambda\mathcal E
  .
  \label{E=}
\end{gather}
Here 
\begin{equation}
  \lambda=
  \left(\int_{-\infty}^{+\infty}|\psi(t,x)|^2\,\d x \right)^{-1}
\end{equation}
is the normalizing constant, which does not depend on $t$ due to Eqs.~\eqref{conserves}, 
\eqref{E-S},
\eqref{E=}. 
Equation for the balance of energy 
\eqref{e-balance} can be rewritten now in the following form:
\begin{gather}
\dot{\rho}=-q', 
\label{rho-balance}
\\
q=\frac{2b\lambda}a \Im 
\left(\psi^\ast\psi'\right).
\end{gather}

Now, the main result of the paper can be formulated as follows:
\begin{proposition} 
  \label{prop-main}
  There is a one-to-one correspondence (a bijection) $(\gamma\{\phii\},\Vv\{\phii\})\leftrightarrow\psi$ between 
\begin{itemize}
  \item
  the set 
  of ordered pairs
  $(\gamma\{\phii\},\ \Vv\{\phii\})$, where $\gamma\{\phii\}$ 
  and $\Vv\{\phii\}$ (strains and particle velocities) are defined by Eqs.~\eqref{Ppsi1} and \eqref{Ppsi2},
  respectively, for any solution $\phii$ of the beam 
  equation~\eqref{beam-gen}
  satisfying arbitrary real valued initial data;
  \item 
  the set 
  of all solutions (wave-functions $\psi$) of the Schrödinger
  equation~\eqref{Spsi} with arbitrary complex valued initial data.
\end{itemize}
This bijection is given by Eq.~\eqref{icPsi}. 
The mechanical energy density 
$\mathcal E\{\phii\}$~\eqref{E-BE}
and the 
probability density $\rho\{\psi\}$~\eqref{E-S} for the corresponding smooth enough solutions $\phii$ and 
$\psi$ are related by
Eq.~\eqref{E=}. The transport equations for $\mathcal E$ and $\rho$ are Eqs.~\eqref{e-balance}
and \eqref{rho-balance}, respectively.
\end{proposition} 

\section{Conclusion}

Though this note discusses equations used in both classical and quantum mechanics, it is written by 
mechanicians mostly for mechanicians. The main result formulated in the form of
Proposition~\ref{prop-main} brings mechanicians the following goals:

\begin{enumerate}
\item 
The possibility to reduce the order of the equation we need to solve dealing with a beam, at the cost
of we find only the strain and the particle velocity, not the displacements. 
\item 
Energetic identity \eqref{E=} provides a clear, intuitive representation of how the quantum probability
density propagates. For example, the 1D problem concerning
the free propagation of the gaussian wave packet \cite{Darwin1927} gets
a simple mechanical interpretation. 
\end{enumerate}
Note that the requirement concerning the initial data for the beam equation to be real valued is
essential for the discussed bijection. In the case of complex valued initial
data for the beam equation, the consideration of both complex conjugate
Schrödinger equations is required, see Proposition~\ref{prop:S2}.

We suppose that the suggested in our note bijection between quantum
and mechanical systems is simpler than the one discussed in 
\cite{Engstrom2023} and involves only one beam.
The generalization of the obtained results to the cases of a non-zero external
potential and/or to 2D-3D equations is possible, but not straightforward. In the 3D case, apparently, 
the particular case of the Cosserat pseudo-continuum 
\cite{Aero1961,Mindlin1962,Grekova2020}
can be taken as an equivalent mechanical system. 
According to energetic identity \eqref{E=}, we expect that two corresponding equations 
(the Euler-Bernoulli and the Schrödinger ones)
are also equivalent in the framework of the recently suggested conception of energy dynamics 
\cite{Krivtsov2022,Baimova2023,Kuzkin2023}.

\section*{Declaration of competing interest}
The authors declare that they have no known competing financial interests or personal relationships that could have appeared
to influence the work reported in this paper.

\section*{Data availability}
No data was used for the research described in the article.

\section*{Acknowledgement}
The authors are grateful to 
A.A.~Sokolov who attracted our attention to this problem and to
E.F.~Grekova, E.A.~Ivanova, V.A.~Kuzkin, Yu.A.~Mochalova, S.A.~Rukolaine 
for discussions.

\section*{Funding}
The research is supported by the Ministry of Science and Higher Education of
the Russian Federation (project 124041500007-4).

\printcredits

\bibliographystyle{elsarticle-num-names}
\bibliography{bib/all}

\begin{thebibliography}{23}
\expandafter\ifx\csname natexlab\endcsname\relax\def\natexlab#1{#1}\fi
\providecommand{\url}[1]{\texttt{#1}}
\providecommand{\href}[2]{#2}
\providecommand{\path}[1]{#1}
\providecommand{\DOIprefix}{doi:}
\providecommand{\ArXivprefix}{arXiv:}
\providecommand{\URLprefix}{URL: }
\providecommand{\Pubmedprefix}{pmid:}
\providecommand{\doi}[1]{\href{http://dx.doi.org/#1}{\path{#1}}}
\providecommand{\Pubmed}[1]{\href{pmid:#1}{\path{#1}}}
\providecommand{\bibinfo}[2]{#2}
\ifx\xfnm\relax \def\xfnm[#1]{\unskip,\space#1}\fi
\bibitem[{Kiselev(2007)}]{Kiselev2007}
\bibinfo{author}{A.~P. Kiselev},
\newblock \bibinfo{title}{Localized light waves: Paraxial and exact solutions of the wave equation (a review)},
\newblock \bibinfo{journal}{Optics and Spectroscopy} \bibinfo{volume}{102} (\bibinfo{year}{2007}) \bibinfo{pages}{603--622}. \DOIprefix\doi{10.1134/S0030400X07040200}.
\bibitem[{Perel et~al.(2005)Perel, Kaplunov, and Rogerson}]{Perel2005}
\bibinfo{author}{M.~V. Perel}, \bibinfo{author}{J.~D. Kaplunov}, \bibinfo{author}{G.~A. Rogerson},
\newblock \bibinfo{title}{An asymptotic theory for internal reflection in weakly inhomogeneous elastic waveguides},
\newblock \bibinfo{journal}{Wave Motion} \bibinfo{volume}{41} (\bibinfo{year}{2005}) \bibinfo{pages}{95--108}. \DOIprefix\doi{10.1016/j.wavemoti.2004.06.001}.
\bibitem[{Schrödinger(1926)}]{SchroedingerAdP4-1926}
\bibinfo{author}{E.~Schrödinger},
\newblock \bibinfo{title}{Quantisierung als {Eigenwertproblem}},
\newblock \bibinfo{journal}{Annalen der Physik} \bibinfo{volume}{386} (\bibinfo{year}{1926}) \bibinfo{pages}{109--139}. \DOIprefix\doi{10.1002/andp.19263861802}.
\bibitem[{Schrödinger(1982)}]{Schroedinger1982}
\bibinfo{author}{E.~Schrödinger}, \bibinfo{title}{Collected Papers on Wave Mechanics}, \bibinfo{publisher}{Chelsea}, \bibinfo{address}{New York}, \bibinfo{year}{1982}.
\bibitem[{Engstrom(2023)}]{Engstrom2023}
\bibinfo{author}{T.~A. Engstrom},
\newblock \bibinfo{title}{Dynamics of certain {E}uler-{B}ernoulli rods and rings from a minimal coupling quantum isomorphism},
\newblock \bibinfo{journal}{Physical Review E} \bibinfo{volume}{107} (\bibinfo{year}{2023}) \bibinfo{pages}{065005}. \DOIprefix\doi{10.1103/PhysRevE.107.065005}.
\bibitem[{Makris(2023)}]{Makris2023}
\bibinfo{author}{N.~Makris},
\newblock \bibinfo{title}{Revisiting {Schr{\"{o}}dinger{\textquoteright}s} fourth-order, real-valued wave equation and the implication from the resulting energy levels},
\newblock \bibinfo{journal}{Royal Society Open Science} \bibinfo{volume}{10} (\bibinfo{year}{2023}). \DOIprefix\doi{10.1098/rsos.230793}.
\bibitem[{Volovich(2024)}]{Volovich2024}
\bibinfo{author}{I.~Volovich},
\newblock \bibinfo{title}{On the equivalence between the {Schrodinger} equation in quantum mechanics and the {Euler}-{Bernoulli} equation in elasticity theory},
\newblock \bibinfo{journal}{arXiv preprint 2411.03261}  (\bibinfo{year}{2024}). \DOIprefix\doi{10.48550/arXiv.2411.03261}.
\bibitem[{Schr{\"o}dinger(1914)}]{schrodinger1914dynamik}
\bibinfo{author}{E.~Schr{\"o}dinger},
\newblock \bibinfo{title}{{Z}ur {D}ynamik elastisch gekoppelter {P}unktsysteme},
\newblock \bibinfo{journal}{Annalen der Physik} \bibinfo{volume}{349} (\bibinfo{year}{1914}) \bibinfo{pages}{916--934}. \DOIprefix\doi{10.1002/andp.19143491405}.
\bibitem[{M\"uhlich et~al.(2020)M\"uhlich, Abali, and {d}ell'Isola}]{Muehlich2020}
\bibinfo{author}{U.~M\"uhlich}, \bibinfo{author}{B.~E. Abali}, \bibinfo{author}{F.~{d}ell'Isola},
\newblock \bibinfo{title}{Commented translation of {E}rwin {S}chr\"odinger's paper `{O}n the dynamics of elastically coupled point systems' ({Z}ur {D}ynamik elastisch gekoppelter {P}unktsysteme)},
\newblock \bibinfo{journal}{Mathematics and Mechanics of Solids} \bibinfo{volume}{26} (\bibinfo{year}{2020}) \bibinfo{pages}{133--147}. \DOIprefix\doi{10.1177/1081286520942955}.
\bibitem[{Messiah(1961)}]{Messiah1-1961}
\bibinfo{author}{A.~Messiah}, \bibinfo{title}{Quantum Mechanics}, volume~\bibinfo{volume}{1}, \bibinfo{publisher}{North-Holland}, \bibinfo{address}{Amsterdam}, \bibinfo{year}{1961}.
\bibitem[{Weaver et~al.(1990)Weaver, Timoshenko, and Young}]{Weaver1990}
\bibinfo{author}{W.~Weaver, Jr.}, \bibinfo{author}{S.~P. Timoshenko}, \bibinfo{author}{D.~H. Young}, \bibinfo{title}{Vibration Problems in Engineering}, \bibinfo{publisher}{Wiley}, \bibinfo{address}{New York}, \bibinfo{year}{1990}.
\bibitem[{Orsingher and D’Ovidio(2011)}]{Orsingher2011}
\bibinfo{author}{E.~Orsingher}, \bibinfo{author}{M.~D’Ovidio},
\newblock \bibinfo{title}{Vibrations and fractional vibrations of rods, plates and {F}resnel pseudo-processes},
\newblock \bibinfo{journal}{Journal of Statistical Physics} \bibinfo{volume}{145} (\bibinfo{year}{2011}) \bibinfo{pages}{143--174}. \DOIprefix\doi{10.1007/s10955-011-0309-5}.
\bibitem[{Korikov et~al.(2021)Korikov, Plamenevskii, and Sarafanov}]{KorikovSch2021}
\bibinfo{author}{D.~Korikov}, \bibinfo{author}{B.~Plamenevskii}, \bibinfo{author}{O.~Sarafanov}, \bibinfo{title}{Schroedinger and Germain-Lagrange Equations in a Domain with Corners}, volume \bibinfo{volume}{284} of \textit{\bibinfo{series}{Operator Theory: Advances and Applications}}, \bibinfo{publisher}{Springer}, \bibinfo{year}{2021}, pp. \bibinfo{pages}{245--293}. \DOIprefix\doi{10.1007/978-3-030-65372-9_6}.
\bibitem[{Altenbach et~al.(2013)Altenbach, Bîrsan, and Eremeyev}]{Altenbach2013}
\bibinfo{author}{H.~Altenbach}, \bibinfo{author}{M.~Bîrsan}, \bibinfo{author}{V.~A. Eremeyev},
\newblock \bibinfo{title}{Cosserat-type rods},
\newblock in: \bibinfo{editor}{H.~Altenbach}, \bibinfo{editor}{V.~A. Eremeyev} (Eds.), \bibinfo{booktitle}{Generalized Continua from the Theory to Engineering Applications}, volume \bibinfo{volume}{541} of \textit{\bibinfo{series}{CISM International Centre for Mechanical Sciences}}, \bibinfo{publisher}{Springer}, \bibinfo{address}{Vienna}, \bibinfo{year}{2013}, pp. \bibinfo{pages}{179--248}. \DOIprefix\doi{10.1007/978-3-7091-1371-4_4}.
\bibitem[{Zhilin(2007)}]{Zhilin2007}
\bibinfo{author}{P.~A. Zhilin}, \bibinfo{title}{Prikladnaya mekhanika. {T}eoriya tonkikh uprugikh sterzhney [{A}pplied mechanics. {T}heory of thin elastic rods]}, \bibinfo{publisher}{St. Petersburg Polytechnic University (SPbPU)}, \bibinfo{address}{St. Petersburg}, \bibinfo{year}{2007}. \bibinfo{note}{In Russian}.
\bibitem[{Vladimirov(1971)}]{Vladimirov1971}
\bibinfo{author}{V.~S. Vladimirov}, \bibinfo{title}{Equations of Mathematical Physics}, \bibinfo{publisher}{Marcel Dekker}, \bibinfo{address}{New York}, \bibinfo{year}{1971}.
\bibitem[{Darwin(1927)}]{Darwin1927}
\bibinfo{author}{C.~G. Darwin},
\newblock \bibinfo{title}{Free motion in the wave mechanics},
\newblock \bibinfo{journal}{Proceedings of the Royal Society of London. Series A} \bibinfo{volume}{117} (\bibinfo{year}{1927}) \bibinfo{pages}{258--293}. \DOIprefix\doi{10.1098/rspa.1927.0179}.
\bibitem[{Aero and Kuvshinskii(1961)}]{Aero1961}
\bibinfo{author}{E.~L. Aero}, \bibinfo{author}{E.~V. Kuvshinskii},
\newblock \bibinfo{title}{Fundamental equations of the theory of elastic media with rotationally interacting particles},
\newblock \bibinfo{journal}{Soviet Physics, Solid State} \bibinfo{volume}{2} (\bibinfo{year}{1961}) \bibinfo{pages}{1272--1281}.
\bibitem[{Mindlin and Tiersten(1962)}]{Mindlin1962}
\bibinfo{author}{R.~D. Mindlin}, \bibinfo{author}{H.~F. Tiersten},
\newblock \bibinfo{title}{Effects of couple-stresses in linear elasticity},
\newblock \bibinfo{journal}{Archive for Rational Mechanics and Analysis} \bibinfo{volume}{11} (\bibinfo{year}{1962}) \bibinfo{pages}{415--448}. \DOIprefix\doi{10.1007/BF00253946}.
\bibitem[{Grekova et~al.(2020)Grekova, Porubov, and dell{\textquoteright}Isola}]{Grekova2020}
\bibinfo{author}{E.~F. Grekova}, \bibinfo{author}{A.~V. Porubov}, \bibinfo{author}{F.~dell{\textquoteright}Isola},
\newblock \bibinfo{title}{Reduced linear constrained elastic and viscoelastic homogeneous {Cosserat} media as acoustic metamaterials},
\newblock \bibinfo{journal}{Symmetry} \bibinfo{volume}{12} (\bibinfo{year}{2020}) \bibinfo{pages}{521}. \DOIprefix\doi{10.3390/sym12040521}.
\bibitem[{Krivtsov(2022)}]{Krivtsov2022}
\bibinfo{author}{A.~M. Krivtsov},
\newblock \bibinfo{title}{Dynamics of matter and energy},
\newblock \bibinfo{journal}{ZAMM --- Journal of Applied Mathematics and Mechanics} \bibinfo{volume}{103} (\bibinfo{year}{2022}). \DOIprefix\doi{10.1002/zamm.202100496}.
\bibitem[{Baimova et~al.(2023)Baimova, Bessonov, and Krivtsov}]{Baimova2023}
\bibinfo{author}{J.~A. Baimova}, \bibinfo{author}{N.~M. Bessonov}, \bibinfo{author}{A.~M. Krivtsov},
\newblock \bibinfo{title}{Motion of localized disturbances in scalar harmonic lattices},
\newblock \bibinfo{journal}{Physical Review E} \bibinfo{volume}{107} (\bibinfo{year}{2023}) \bibinfo{pages}{065002}. \DOIprefix\doi{10.1103/PhysRevE.107.065002}.
\bibitem[{Kuzkin(2023)}]{Kuzkin2023}
\bibinfo{author}{V.~A. Kuzkin},
\newblock \bibinfo{title}{Acoustic transparency of the chain-chain interface},
\newblock \bibinfo{journal}{Physical Review E} \bibinfo{volume}{107} (\bibinfo{year}{2023}) \bibinfo{pages}{065004}. \DOIprefix\doi{10.1103/PhysRevE.107.065004}.

\end{thebibliography}

\end{document}